\documentclass{article}
\usepackage[numbers]{natbib}
\usepackage {arxiv}

\usepackage{cite}
\usepackage{floatrow}
\usepackage{amsmath,amssymb,amsfonts}
\usepackage{algorithm}
\usepackage{algorithmic}
\usepackage{textcomp}
\usepackage{xcolor}
\usepackage{float}
\usepackage{booktabs}
\usepackage{multirow}
\usepackage{enumitem}
\usepackage{setspace}
\usepackage{graphicx}
\usepackage{array}
\usepackage{amsmath}
\usepackage{amssymb}
\usepackage{epsfig}
\usepackage{epstopdf}
\usepackage{subfigure}
\usepackage{comment}
\usepackage{tabularx}
\usepackage{threeparttable}
\usepackage{adjustbox}
\usepackage{soul}
\usepackage{dsfont}
\usepackage[hyphens]{url}
\usepackage{hyperref} 

\makeatletter
 \let\old@ps@headings\ps@headings
 \let\old@ps@IEEEtitlepagestyle\ps@IEEEtitlepagestyle
 \def\confheader#1{
 \def\ps@IEEEtitlepagestyle{%
 \old@ps@IEEEtitlepagestyle%
 \def\@oddhead{\strut\hfill#1\hfill\strut}%
 \def\@evenhead{\strut\hfill#1\hfill\strut}%
 }%
 \ps@headings%
 }
 \makeatother

\confheader{
 }
\usepackage[pscoord]{eso-pic}

\def\BibTeX{{\rm B\kern-.05em{\sc i\kern-.025em b}\kern-.08em
    T\kern-.1667em\lower.7ex\hbox{E}\kern-.125emX}}

\title{\LARGE \bf
A Fair Benchmarking of Deep Relational Database Learning Models}

\author{Kazi F. Akhter, ~
        Manar D. Samad\\
Department of Computer Science\\
Tennessee State University\\
Nashville, TN, USA\\
\texttt{manar.samad@outlook.com} \\
\and 
 {\bf Bharath Ajendla}\\
 SAP AI Research and Innovation\\
 SAP Labs\\
 Palo Alto, CA, USA
}

\begin{document}

\maketitle

\begin{abstract}

Relational databases (RDBs) are the primary data infrastructure in many enterprises, yet recent deep learning methods designed for RDBs have been evaluated under inconsistent experimental protocols, making fair comparison difficult. We present one of the first systematic benchmarking studies of recently released deep learning methods for RDBs, evaluating them across five relational databases, with one classification and one regression task for each. We refactor all deep RDB models to allow the full range of experimental procedures to be applied consistently across all methods. Our findings indicate that the relational transformer (RT) approach delivers the strongest overall performance on both classification and regression tasks compared to the state-of-the-art graph-based modeling and learning of RDBs. Even for single-table learning tasks, deep learning methods designed for RDBs outperform the leading tabular foundation model, TabPFN 2.5. Extending learning from a single table (hop = 0) to multiple tables (hop = 1, 2) by connecting neighboring tables in relational databases enhances performance, but the additional benefit from higher hops diminishes as computational overhead grows. Deep RDB learning methods have the potential to challenge state-of-the-art tabular foundation models, especially on large-scale enterprise data. The source code for this benchmarking study is publicly available at~\footnote{\href{https://github.com/mdsamad001/Benchmarking-Deep-Relational-Database-Models.git}%
{https://github.com/mdsamad001/Benchmarking-Deep-Relational-Database-Models.git}}.

\end{abstract}



\section{Introduction}

Relational databases (RDBs) are the backbone of structured data management, where multiple interconnected tables are linked via primary-foreign key relationships. Countless services are performed in e-commerce, healthcare, finance, banking and scientific research by managing data in dozens of tables, millions of rows, and hundreds of columns.~\citep{RELBENCH_Robinson2024, fey2024position}. RDBs are designed and optimized to support complex data queries and management, but they do not take into account the requirements of predictive modeling. Many consequential predictive problems, including customer churn forecasting, clinical risk scoring, supply-chain demand estimation, and recommendation, can benefit from learning RDBs~\citep{fey2024position,fey2024rdl}. However, the sheer volume and breadth of RDB data spread across dozens of interrelated tables and millions of rows pose a unique challenge for developing big-data-driven predictive models. The fundamental challenge in applying conventional machine learning paradigms to RDBs stems from the structural distinctions between ``tabular'' and ``relational'' learning. In a typical tabular structure, each data sample is a self-contained row where samples are assumed to be independent and identically distributed (i.i.d.). RDBs are different from typical tabular data sets in three different ways. First, an RDB includes multiple related tables linked by constraints of the primary-key and foreign-key (PK--FK), which also contribute the necessary predictive signals to learn an entity type~\citep{RT_Ranjan2025}. 
For example, predicting an entity type, such as customers, would benefit from customer-related information spread across other tables that store transactions, orders, and reviews. Second,  RDBs commonly include timestamps and historical records, unlike a data table with static information.   Time information must be properly considered to avoid temporal leakage in predictive modeling tasks~\citep{RELBENCH_Robinson2024}. Third, samples in RDB are inherently dependent, unlike the i.i.d. assumption made in traditional machine learning. For example, two customers can order similar products and services or the same patient appears in multiple follow-up visits, introducing statistical dependencies between samples~\citep{fey2024rdl}. These three distinctions have motivated recent research efforts in introducing deep learning methods that operate directly on RDBs leveraging the relational information across tables. However, deep learning methods for RDBs have been evaluated under varying training configurations and computational budgets against some selective baselines. Therefore, it is unclear whether the performance gain reported for the existing methods is significant compared to other competitive baselines. The tabular learning community has benefited from extensive benchmarking studies~\citep{grinsztajn2022tree, borisov2022survey, shwartz2022tabular, gorishniy2021revisiting}. We argue that recent deep RDB methods have achieved a level of maturity that warrants a first of its kind benchmarking study.

In this work, we present a systematic evaluation of state-of-the-art deep learning methods for relational databases. Our controlled experiments with a common protocol compare across representative architectures that span GNN-based methods, graph transformers, and foundation model approaches. We do not intend to propose a new method, but to inform the community with an empirical characterization of state-of-the-art methods to further guide future research and practical deployment decisions.

\section{Related Work}
\label{sec:related_work}

\subsection{Single-Table Tabular Methods}
\label{sec:rw_tabular}

Deep learning for tabular data has long struggled to demonstrate robust performance compared to traditional machine learning methods~\citep{Survey_Borisov2022, Survey_Grinsztajn2022}. Unfortunately, there is no single best learning algorithm for all tabular datasets due to heterogeneity in domains and tabular structures~\citep{rabbani2025_benchmark}, unlike image and language models. Gradient-boosted decision trees, such as LightGBM, remain one of the dominant methods for tabular prediction tasks due to their robust performance and fast training without the requirement of feature engineering~\citep{Survey_Grinsztajn2022}. 
RealMLP has emerged as a competitive deep learning model for tabular data, demonstrating that improved MLP architectures can achieve performance comparable to gradient-boosted decision trees on large tabular benchmarks~\citep{holzmuller2024realmlp}.
Recently, TabPFN 2.5~\citep{hollmann2025tabpfn} has topped the tabular model leaderboard, which uses a transformer model trained on millions of synthetic tabular datasets. Training examples, such as synthetic data for TabPFN 2.5, are generated by simulating millions of tabular datasets using structural causal models that create diverse relationships between features and targets and include realistic challenges such as noise, nonlinear transformations, and categorical variables.

\begin{figure*}[t]
\centerline{\includegraphics[width=1.0\textwidth]{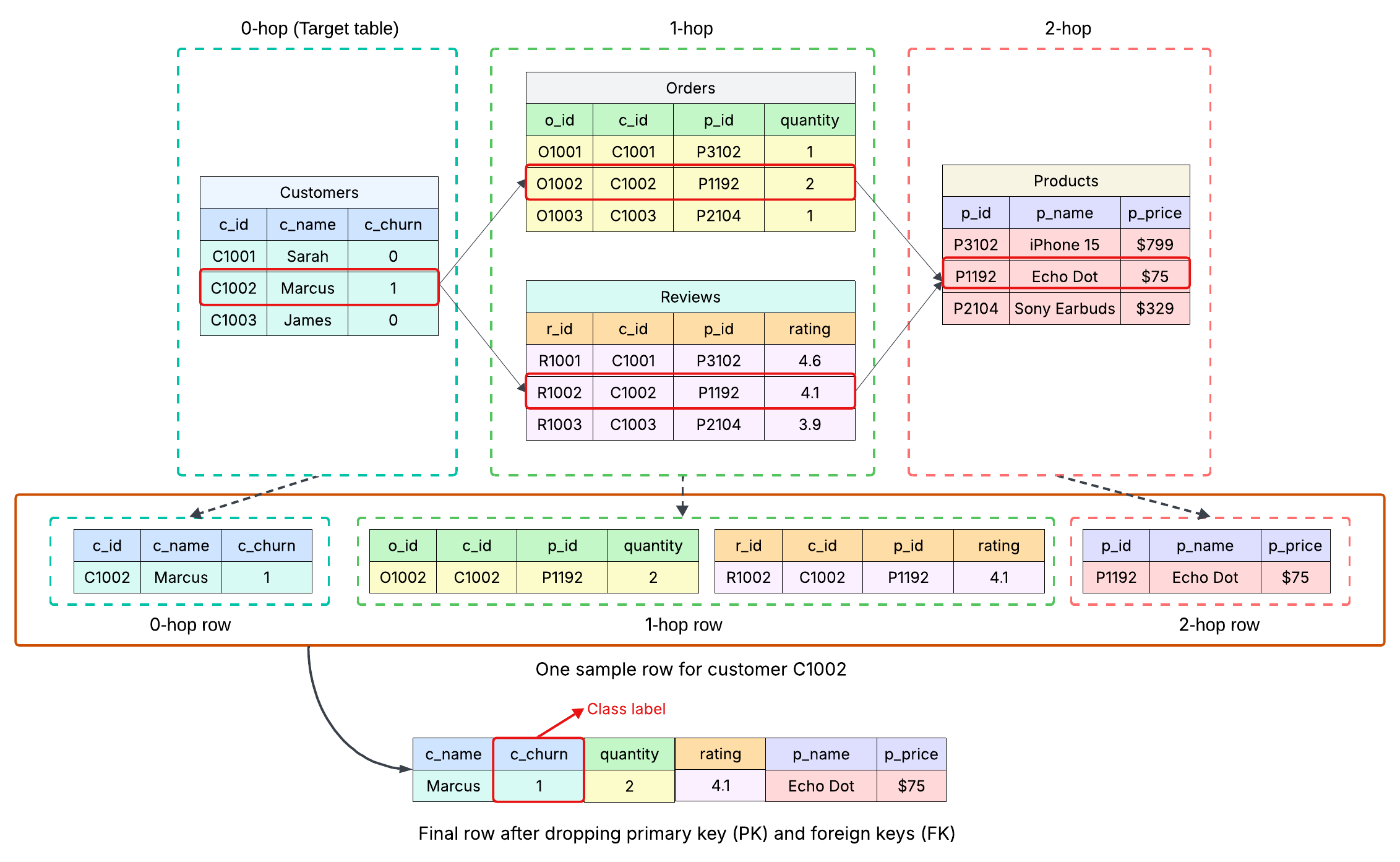}}
\vspace{-10pt}
\caption{Multi-table relationships in a relational database to illustrate hop-based relational neighborhood expansion for a target row $u$ = C1002. Each hop retrieves rows from tables connected via primary--foreign key (PK--FK) links: the \textbf{Customers} table at 0-hop, \textbf{Orders} and \textbf{Reviews} at 1-hop, and \textbf{Products} at 2-hop. The final sample (for 2-hop configuration) concatenates feature columns across all hops, with PK and FK columns removed. The column \texttt{c\_churn} serves as the class label.}
\label{fig-database-hops}
\end{figure*}

The final prediction on the test data is achieved using a single forward pass without any gradient-based finetuning. The recent TabPFN model has shown competitive performance, strongly outperforming traditional machine learning baselines in small-to moderate-sized data sets (around 10,000 samples and 500 features). However, a recent study~\citep{ye2025tabpfnlimitation} also shows that the TabPFN 2.5 model struggles against tree-based models on large-scale data sets. Because relational databases involve millions of records across multiple tables, they may pose a real challenge for TabPFN 2.5.

\begin{table*}[t]
\centering
\caption{Summary of selected RelBench databases and tasks used for benchmarking. Task rows include actual samples relevant to the predictive tasks, which are then split into train, validation, and test rows. Database size is based on the size of the task rows.}
\label{tab:relbench_tasks}
\scalebox{0.95}{
\begin{tabular}{ccccccccc}
\toprule
\textbf{Database} 
& \textbf{Task} 
& \textbf{Task Type} 
& \textbf{\begin{tabular}[c]{@{}c@{}}Database\\Rows\end{tabular}} 
& \textbf{\begin{tabular}[c]{@{}c@{}}Task\\Rows\end{tabular}} 
& \textbf{\begin{tabular}[c]{@{}c@{}}Train\\Rows\end{tabular}} 
& \textbf{\begin{tabular}[c]{@{}c@{}}Validation\\Rows\end{tabular}} 
& \textbf{\begin{tabular}[c]{@{}c@{}}Test\\Rows\end{tabular}} 
& \textbf{\begin{tabular}[c]{@{}c@{}}Sample\\Size\end{tabular}} \\
\midrule
\multirow{2}{*}{rel-amazon} 
  & item-churn    & Classification & 15,000,713 & 2,903,795 & 2,559,264 & 177,689 & 166,842 & Large \\
  & item-ltv      & Regression     & 15,000,713 & 3,052,991 & 2,707,679 & 166,978 & 178,334 & Large \\
\midrule
\multirow{2}{*}{rel-avito} 
  & user-visits   & Classification & 20,679,117 & 152,727   & 86,619    & 29,979  & 36,129  & Medium \\
  & ad-ctr        & Regression     & 20,679,117 & 8,682     & 5,100     & 1,766   & 1,816   & Small \\
\midrule
\multirow{2}{*}{rel-f1} 
  & driver-dnf    & Classification & 74,063     & 12,679    & 11,411    & 566     & 702     & Small \\
  & driver-position & Regression   & 74,063     & 8,712     & 7,453     & 499     & 760     & Small \\
\midrule
\multirow{2}{*}{rel-stack} 
  & user-engagement & Classification & 4,247,264 & 1,534,825 & 1,360,850 & 85,838  & 88,137  & Large \\
  & post-votes    & Regression     & 4,247,264  & 2,771,040 & 2,453,921 & 156,216 & 160,903 & Large \\
\midrule
\multirow{2}{*}{rel-trial} 
  & study-outcome & Classification & 5,434,924  & 13,779    & 11,994    & 960     & 825     & Small \\
  & study-adverse & Regression     & 5,434,924  & 50,029    & 43,335    & 3,596   & 3,098   & Small \\
\bottomrule
\end{tabular}
}
\end{table*}

All learning methods for tabular data assume a single table with self-contained rows and columns. However, many data tables, namely relations, from business, finance, and medical enterprises appear in RDBs, connected by referential integrity, as shown in Figure~\ref{fig-database-hops}. Therefore, a learning task (e.g., classification or regression) related to a single table in an RDB can benefit from incorporating data drawn from its neighboring tables linked through referential keys. The deep learning methods proposed for single tables can be used by joining multiple related tables into a single table. However, the joining process inevitably loses important relational information~\citep{RDNN_You2024, RdbtoGraph_Cvitkovic2020} and generates missing values when the rows between tables do not have one-to-one mappings. Over the past few years, a number of deep architectures have been proposed, tailored to learning the multi-table referential structure of RDBs. Among the proposed architectures, graph neural networks are the most common, where RDBs are modeled using heterogeneous graphs~\citep{fey2024rdl, RELBENCH_Robinson2024, chen2025relgnn}. The attention mechanism on the relational structure is introduced in the graph transformer~\citep{dwivedi2025relgt}. More recently, graph-based~\citep{wang2025griffin} and transformer-based~\citep{RT_Ranjan2025} foundation models for relational databases have been proposed to generalize across different database schemas.

\subsection{Graph-based learning methods}
\label{sec:rw_griffin}
Joining multiple tables into a single table can compromise useful relationships between tables, which can be preserved by modeling multi-table structures as graphs for processing with neural networks. Griffin~\citep{wang2025griffin} models an RDB as a graph with rows as nodes and their relationships (primary-foreign key links between rows across tables) as edges. Griffin models RDBs in graph structures following two steps. Griffin first obtains an initial representation for each row by aggregating its column-level information. In other words, attention between the column-level cell embeddings of each row is used to obtain the corresponding row embedding. The second step in Griffin involves information exchange between rows across tables in RDBs. A message-passing mechanism is applied to update the row embeddings (representations) by propagating messages across connected rows. The message passing follows the edges in a graph representing the primary-foreign key relationship: a row in one table sends its embedding to connected rows in other tables. Each row updates its embedding by aggregating the incoming messages from connected rows. A special feature of Griffin is that it is designed as a foundation model. It includes pretrained text and numerical encoders to handle diverse data types in RDBs. Griffin is pretrained on both single-table and multi-table databases to learn general representations that transfer across databases and tasks, enabling a single pretrained model to be adapted through task-specific fine-tuning for various downstream applications.

\subsection{Transformer-based learning methods}
\label{sec:rw_rt}

The Relational Transformer (RT)~\citep{RT_Ranjan2025} introduces a fine-grained representation of individual tabular cells in RDB. For each table, every cell value is encoded as a vector representation, including the row, column, and table identifiers of the cell. For a given prediction task, the model uses the target row sample whose label is to be predicted and combines all cell embeddings of that row and those from rows in other tables linked by primary-foreign key relationships. The combined cell representations from multiple tables are used to learn a between-cell attention-weighted representation using a transformer. However, sparse attention is maintained by restricting cell pairs in attention computation. Cell pairing is regulated using a fixed binary self-attention mask. The binary mask allows a cell representation to attend to other cells if they belong to the same row (capturing interactions between columns), the same column within a table (capturing interactions between rows), or cells belonging to rows of different tables that are connected via primary-foreign key relationships. All other cell pairs are masked out in computing attention and do not attend to each other.

\subsection{Hybrid graph-transformer learning methods}
\label{sec:rw_dbformer}

DBFormer~\citep{DBformer_pelevska2025} uses a hybrid graph-transformer architecture. Similarly to graph-based learning methods such as Griffin~\citep{wang2025griffin}, the relational structure of the database is represented by rows as nodes and primary-foreign-key relationships as edges. However, DBFormer differs from Griffin in two important ways: 1) how it processes information \emph{within} a table through the attention mechanism and 2) how it exchanges information \emph{between} tables through graph modeling. Within each table, DBFormer~\citep{DBformer_pelevska2025} uses a transformer encoder to learn each row representation using the self-attention mechanisms. The self-attention mechanisms learn attention between columns to capture feature interactions in row representation. To learn between tables, DBFormer connects rows between tables via primary-foreign key (PK-FK) links. However, instead of using message passing via standard GNN aggregation, DBFormer uses cross-attention between the cells (features) of connected rows to learn PK-FK representation.  
This process is repeated across multiple transformer layers , allowing information to propagate across multi-hop relational neighborhoods. One point to note is that DBformer is not designed as a foundation model, unlike its counterparts RT and Griffin models. Therefore, DBformer must be trained on individual RDBs data from scratch.

\begin{table}[t]
\centering
\caption{Comparison of training protocols, neighborhood construction, and hops across relational learning models under their original settings and our controlled benchmarking setup.}
\label{tab:benchamark-protocols}
\scalebox{0.75}{
\begin{tabular}{lcccc}
\toprule
\textbf{Factors} 
& \textbf{DBFormer} 
& \textbf{Griffin} 
& \textbf{RT} 
& \textbf{Our Benchmarking} \\
\midrule
Model Selection       & None           & Early stopping                & None        & Best model on Val.                           \\
Train Epoch          & 1000           & 50                  & None        & 100                                          \\
0-Hop (single table)                & N/A            & Yes                 & N/A         & Yes                            \\
1-Hop                & N/A            & N/A                 & N/A         & Yes                            \\
2-Hop                & N/A            & Yes                 & N/A         & Yes                            \\
Databases            & CTU repository & RelBench, 4DBInfer  & RelBench   & RelBench                       \\
Neighborhood control & N/A            & N/A      & N/A         & 1-Hop, 2-Hop \\
Embedding size       & 64             & 256                 & 512         & 256                            \\
Learning rate        & 1e-5           & 3e-4                & 1e-4        & 1e-4                           \\
Batch size           & 16             & 256                 & 256         & 256                            \\
\bottomrule
\end{tabular}}
\end{table}


\subsection{Need for systematic benchmarking}
\label{sec:rw_benchmarking}

Existing deep RDB methods are evaluated using varying experimental protocols, computational budgets, and against selective baselines~\citep{DBformer_pelevska2025, wang2025griffin, RT_Ranjan2025}, which makes a fair comparison of current models challenging. Deep RDB models learn from multiple tables by expanding the relational neighborhood of a target row outward through primary-foreign key links, as illustrated in Figure~\ref{fig-database-hops}. Practical factors beyond predictive performance, such as training time and memory usage, are critical to processing large RDBs with up to millions of samples. However, computational time and complexity are rarely reported in the deep RDB learning literature. For example, the training time reported in RT~\citep{RT_Ranjan2025} is for a specific large-scale hardware configuration (8 GPUs/320GB of memory), which is not directly comparable to the hardware used in other studies. Moreover, the performance of a learning algorithm can vary significantly for different databases and tasks, since there is no universal best performer for all data problems~\citep{rabbani2025attention}. Therefore, the selection of databases and learning tasks is a critical step in evaluating the learning algorithms. For example, the study-outcome classification task in the \texttt{rel-trial} RDB shows superior performance on the pretrained Griffin~\citep{wang2025griffin} model compared to its non-pretrained counterpart. The opposite performance trend is observed for the user-clicks classification task in the \texttt{rel-avito} RDB. Consequently, a standardized benchmarking protocol is needed to allow a fair and independent assessment of the deep RDB models proposed in the literature. The tabular learning community has benefited substantially from this kind of controlled evaluation~\citep{Survey_Grinsztajn2022, Survey_Borisov2022}. Relational deep learning, despite its rapidly expanding model landscape, lacks a comparable body of research. Our work aims to fill this gap by evaluating all models in a unified protocol with matched computational budgets, consistent data splits, and standardized evaluation metrics.


\section {Methodology}

\begin{table*}[t]
\caption{AUROC scores of RelBench classification tasks. Average rank is obtained for each hop setting (0, 1, 2) across five classification tasks and five classification methods.}
\label{tab:classification_results}
\centering
\scalebox{0.9}{
\begin{tabular}{llccccccccccccccc}
\toprule
& & \multicolumn{3}{c}{LightGBM} 
& \multicolumn{3}{c}{TabPFN 2.5}
& \multicolumn{3}{c}{DBFormer}
& \multicolumn{3}{c}{Griffin}
& \multicolumn{3}{c}{RT} \\
\cmidrule(lr){3-5} \cmidrule(lr){6-8} \cmidrule(lr){9-11} \cmidrule(lr){12-14} \cmidrule(lr){15-17}

\multicolumn{2}{c}{} 
& 0 & 1 & 2 
& 0 & 1 & 2 
& 0 & 1 & 2 
& 0 & 1 & 2 
& 0 & 1 & 2 \\
\midrule

Database & Task \\

\midrule
rel-amazon & item-churn
& 67.7 & 68.1 & 68.6
& 59.0 & 67.0 & 67.7
& 54.0 & 52.4 & 71.8
& 73.1 & 73.2 & 73.2
& 81.6 & 73.6 & 82.2 \\

rel-avito & user-visits
& 53.8 & 58.5 & 60.8
& 53.0 & 64.5 & 59.6
& 52.6 & 59.9 & 60.7
& 50.3 & 62.2 & 63.3
& 64.3 & 63.9 & 65.4 \\

rel-f1 & driver-dnf
& 66.2 & 66.5 & 72.1
& 56.7 & 70.0 & 70.8
& 47.8 & 71.2 & 74.1
& 54.5 & 71.9 & 73.9
& 76.7 & 74.0 & 74.5 \\

rel-stack & user-engagement
& 79.1 & 81.5 & 82.2
& 62.8 & 82.5 & 82.9
& 62.3 & 69.4 & 77.2
& 49.0 & 83.0 & 83.3
& 89.1 & 88.9 & 89.4 \\

rel-trial & study-outcome
& 72.1 & 62.3 & 65.4
& 71.9 & 67.2 & 68.4
& 53.8 & 49.8 & 50.2
& 67.0 & 67.9 & 66.9
& 65.6 & 69.4 & 70.2 \\

\midrule
\multicolumn{2}{c}{Average rank under each hop}
& 2.0 & 4.2 & 3.8
& 3.2 & 3.0 & 4.0
& 4.6 & 4.4 & 3.8
& 3.6 & 2.2 & 2.4
& 1.6 & 1.2 & 1.0 \\
\bottomrule
\end{tabular}
}
\end{table*}

\begin{table*}[t]
\caption{RMSE scores of RelBench regression tasks. Average rank is obtained for each hop setting (0, 1, 2) across five regression tasks and five regression methods.}
\label{tab:regression_results}
\centering
\scalebox{0.9}{
\begin{tabular}{llccccccccccccccc}
\toprule
& & \multicolumn{3}{c}{LightGBM} 
& \multicolumn{3}{c}{TabPFN 2.5}
& \multicolumn{3}{c}{DBFormer}
& \multicolumn{3}{c}{Griffin}
& \multicolumn{3}{c}{RT} \\
\cmidrule(lr){3-5} \cmidrule(lr){6-8} \cmidrule(lr){9-11} \cmidrule(lr){12-14} \cmidrule(lr){15-17}
\multicolumn{2}{c}{Hop} 
& 0 & 1 & 2 
& 0 & 1 & 2 
& 0 & 1 & 2 
& 0 & 1 & 2 
& 0 & 1 & 2 \\
\midrule
\multicolumn{1}{c}{Database} & \multicolumn{1}{c}{Task} \\
\midrule
rel-amazon & item-ltv
& 8.22 & 8.24 & 8.23
& 6.60 & 6.49 & 6.50
& 56.10 & 56.03 & 55.14
& 8.17 & 7.97 & 7.83
& 0.51 & 0.47 & 0.45 \\
rel-avito & ad-ctr
& 3.47 & 3.16 & 3.00
& 3.74 & 3.00 & 2.95
& 0.28 & 0.30 & 0.33
& 0.85 & 0.82 & 0.82
& 1.06 & 1.15 & 1.14 \\
rel-f1 & driver-position
& 6.02 & 5.59 & 5.55
& 5.39 & 4.55 & 4.60
& 6.91 & 11.69 & 11.69
& 0.73 & 0.68 & 0.68
& 0.58 & 0.61 & 0.58 \\
rel-stack & post-votes
& 0.33 & 0.39 & 0.36
& 0.37 & 0.35 & 0.35
& 0.37 & 0.36 & 0.36
& 1.28 & 0.94 & 0.91
& 0.60 & 0.61 & 0.60 \\
rel-trial & study-adverse
& 7.26 & 7.34 & 7.22
& 6.64 & 6.32 & 5.66
& 15.82 & 15.77 & 15.79
& 2.26 & 1.12 & 2.26
& 0.65 & 0.61 & 0.54 \\
\midrule
\multicolumn{2}{c}{Average rank under each hop}
& 3.40 & 3.80 & 3.90
& 3.10 & 2.40 & 2.60
& 3.70 & 3.60 & 3.70
& 2.80 & 2.80 & 2.80
& 2.00 & 2.40 & 2.00 \\
\bottomrule
\end{tabular}
}
\end{table*}

\subsection{Customizing for benchmarking}

Individual deep RDB methods have been implemented using distinct protocols, each requiring adjustments to ensure fair benchmarking in a controlled setting. 
We have developed a unified benchmarking protocol that standardizes data representation, relational context construction, and evaluation for all methods, as shown in Table \ref{tab:benchamark-protocols}.

\subsubsection{Relational mapping in hops}

First, recent studies differ in how they manage and exploit relationships between tables in an RDB. Relational databases store data in multiple connected tables, where relationships between tables (via primary–foreign keys) allow information to be shared across them. Relational models, such as RT~\citep{relationaltransformer2025}, Griffin~\citep{wang2025griffin}, and DBFormer~\citep{DBformer_pelevska2025}, are designed to use connections between tables. However, these methods rely on predefined sampling strategies, e.g., randomly selecting a limited number of rows from neighboring tables. There is no clear control mechanism for how far the referential chain of tables in an RDB is considered when sampling rows. To address this inconsistency, we introduce a hop-based formulation, where a hop represents a one-step referential connection (PK-FK) between two neighboring tables. For a target row or table, the 1-hop neighborhood consists of rows of tables directly linked to the target through a PK-FK relationship. The 2-hop neighborhood consists of rows from tables in the second neighborhood that are connected to the first neighborhood but not directly to the target, as shown in Figure~\ref{fig-database-hops}. The hop setting can be used to control how deeply the model retrieves information from an RDB for a predictive task. By the definition of hop, a hop=0 configuration uses only the target table without including rows from the neighborhood tables. For a given target row \(u\), we define its relational neighborhood using hop-based expansion, where \(\mathcal{N}_h(u)\) denotes the set of rows that belong to the \(h\)-hop neighborhood of \(u\), with \(h \in \{0,1,2\}\) in this study.

The 0-hop neighborhood contains only the target row:
\begin{equation}
\mathcal{N}_0(u) = \{u\}
\end{equation}

The 1-hop neighborhood consists of all rows directly connected to \(u\):
\begin{equation}
\mathcal{N}_1(u) = \{ v \mid v \text{ is directly linked to } u \}
\end{equation}

The 2-hop neighborhood represents rows that are reached by following two consecutive primary--foreign key connections starting from the target row (\emph{u}).
\begin{equation}
\mathcal{N}_2(u) = \{ w \mid w \text{ is linked to some } v \in \mathcal{N}_1(u) \}
\end{equation}

In general, the relational context of \(u\) up to hop \(H\) is:
\begin{equation}
\mathcal{N}_{\le H}(u) = \bigcup_{h=0}^{H} \mathcal{N}_h(u)
\end{equation}
which includes all rows reachable within \(H\) relational steps. This neighborhood is constructed iteratively by expanding the target row through PK-FK connections for \(H\) steps and restricting expansion beyond \(H\) hops.

\subsubsection{Data encoding mechanism}
Second, deep RDB methods use different data encoding methods to convert and combine a heterogeneous feature space (numerical, categorical, or ordinal) of RDB tables into an effective representation vector. For example, DBFormer uses GloVe embeddings~\citep{pennington2014_glove}, Griffin uses Nomic Embed~\citep{nussbaum2024_nomic}, and RT uses MiniLM-v2~\citep{wang2021_minilmv2}.
These encoders differ substantially in their embedding dimensionality and semantic capabilities. The choice of data encoder can influence the performance of deep RDB models. For example, GloVe embeddings are outdated because they do not provide context-aware embeddings, unlike transformer-based encoders. The Nomic Embed encoder contains roughly 137 million parameters and outputs 768-dimensional embeddings, whereas MiniLM-v2 has about 33 million parameters and generates 384-dimensional embeddings—approximately one-quarter of the parameters and half the embedding dimensionality. Since our benchmarking protocol involves training each model multiple times on several large RDBs, tasks, and hop settings, using a lighter encoder significantly reduces computational cost. Therefore, to enable a fair comparison, we employ MiniLM-v2~\citep{wang2021_minilmv2} for all models, which generates 384-dimensional dense embeddings. We use MiniLM-v2~\citep{wang2021_minilmv2}, particularly for encoding categorical and textual features, from SentenceTransformers~\citep{reimers2019_sentencetransformers}. MiniLM-v2 is a compact transformer model obtained via knowledge distillation from larger pretrained model, BERT. Specifically, it learns to mimic the self-attention behavior of the large BERT model while using significantly fewer parameters, resulting in a lightweight yet high-quality contextual encoder.

\begin{figure}[t]
\centering
\subfigure[Classification] { \includegraphics[trim=0.0 0 0 0cm, clip, width=0.45\textwidth] {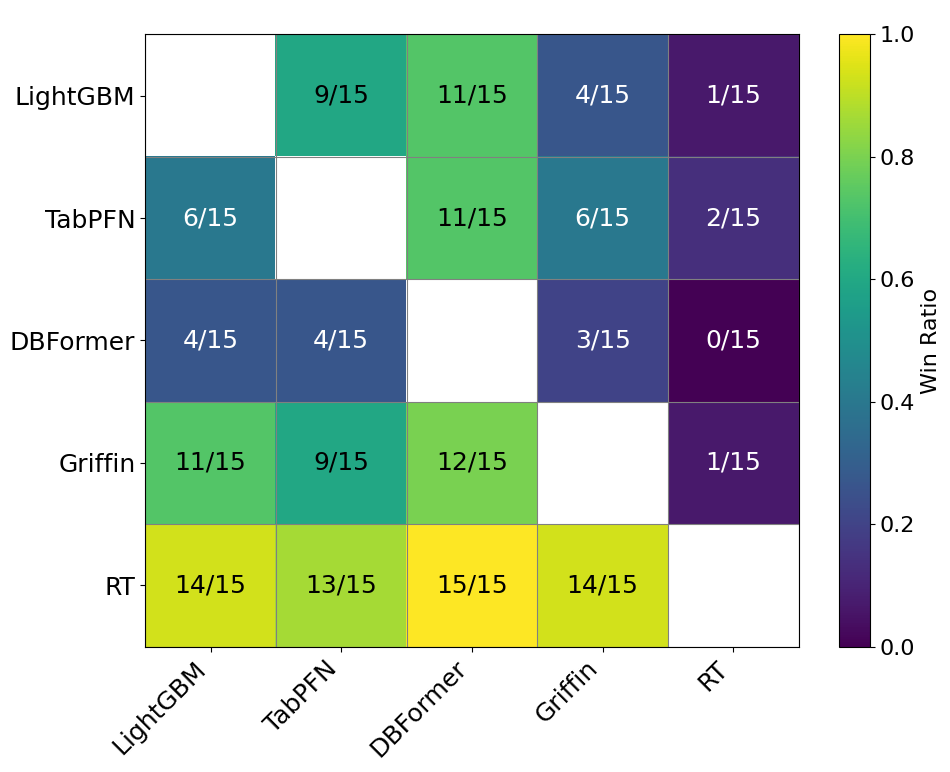}
\label{fig:win_matrix_classification}
}
\hspace{-10pt}
\subfigure[Regression] { \includegraphics[trim=0.0 0 0 0cm, clip, width=0.45\textwidth] {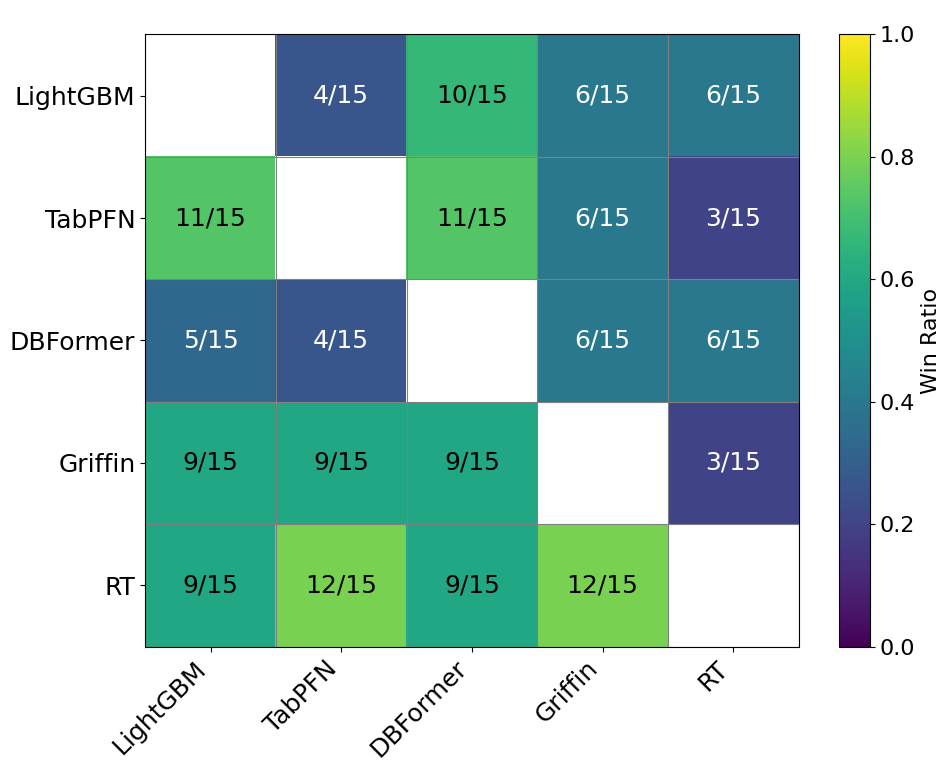}
\label{fig:win_matrix_regression}
}
\caption{Pairwise model comparison using the win–loss matrix. (a) For classification tasks using AUROC scores and (b) for regression tasks using RMSE scores. Each row model shows the number of experimental cases (out of 15) it is superior to the models in the columns. A total of 15 experimental cases consist of three different hops and five databases. 
}
\vspace{-10pt}
\label{win_matrix}
\end{figure}

Third, different encoding strategies for numerical features in RDBs have been adopted in the literature. For example, DBFormer uses a simple linear layer to convert each numeric value into a vector, while RT uses a similar linear transformation to map each value to the hidden dimension of the model. In short, both DBFormer and RT use a simple linear layer to project each numerical value directly into their respective model. In comparison, Griffin employs a more advanced strategy. It begins by normalizing numerical values with a quantile transformer. This transformer, provided by scikit-learn~\citep{pedregosa2011_scikit}, applies a statistical procedure that maps numerical inputs onto a normal distribution. After this step, the normalized values are fed into a pretrained encoder–decoder pair, both realized as Multi-Layer Perceptrons (MLPs).

\subsubsection{Customizing non-relational baselines}
In this study, we consider state-of-the-art non-relational baseline methods for deep learning on tabular data. These methods need to be extended for multi-hop learning to ensure a fair comparison. For TabPFN 2.5~\citep{hollmann2025tabpfn} and LightGBM, we convert multi-hop relational information into tabular features by aggregating data from connected tables, similar to how RelBench~\citep{RELBENCH_Robinson2024} compares traditional machine learning models against its proposed GNN-based Relational Deep Learning (RDL) model. However, unlike RelBench~\citep{RELBENCH_Robinson2024}, we
control the relational context by limiting the hop distance (\emph{H}) and enforcing temporal consistency to avoid information leakage.

\subsection{Effect of multi-hop deep RDB learning}

Deep RDB learning uses a task-specific target table. A learning task that uses only the target table is defined as having hop = 0, meaning that no information from neighboring tables connected to the target is used. As shown in Figure~\ref{fig-database-hops}, the hop = 1 scenario incorporates the tables directly adjacent to the target table. Similarly, in the hop=2 scenario, the network is expanded to include the second-tier tables that are directly linked to the adjacent tables. The computational complexity increases as we expand the network of tables from hop =0 to hop=2. The effect of network expansion on overall learning performance has not been well studied in the deep RDB literature. 

\section{Results}

\subsection {Relational databases and computing platform}

In recent work on deep learning for RDB~\citep{wang2025griffin, relationaltransformer2025}, RelBench appears as the primary benchmarking repository~\citep{RELBENCH_Robinson2024}.  We use five databases from RelBench, assigning two tasks to each: one classification task and one regression task. The selected databases and tasks are summarized in Table~\ref{tab:relbench_tasks}. For each database, Table~\ref{tab:relbench_tasks} reports the total number of rows in all tables (database rows), the number of samples available for the prediction task at hop = 0 (task rows), and the number of training, validation and test samples derived from the task rows following the RelBench split. The tasks are further grouped into three categories based on the scale of the task rows.  Small tasks contain fewer than 100,000 rows, medium tasks contain fewer than 1 million task rows, and large tasks contain at least 1 million task rows.

Deep relational models (DBFormer, Griffin and RT) are trained and evaluated on an NVIDIA DGX Spark GB10, a personalized AI supercomputer with a 20-core ARM CPU, 128 GB of unified memory, and Ubuntu 24.04 LTS operating system. The baseline methods for learning tabular data (LightGBM and TabPFN 2.5) are run on an Ubuntu 22.04 workstation with a 13th Gen Intel Core i9-13900F 24-core processor, 64, GB of RAM, and an NVIDIA GeForce RTX 4090 GPU with 24 GB of VRAM.

\subsection {Classification performance}
\label{classification-performance}

The single table classification task (hop = 0) reveals that RT is overall the best method (average rank 1.6) followed by LightGBM. Griffin places third, while TabPFN 2.5—arguably the best-reported method for learning tabular data, ranks fourth. The underperformance of TabPFN 2.5 may be surprising, but note that relational databases typically contain far more samples than traditional tabular datasets. It is worth mentioning that TabPFN 2.5 claims to outperform other state-of-the-art approaches on small datasets, specifically those with fewer than 10,000 samples~\citep{ye2025tabpfnlimitation}. DBformer appears as the worst of all methods. Furthermore, the DBformer authors have benchmarked it only on databases from the CTU repository, underscoring the need for this benchmarking study. An interesting exception is the \texttt{rel-trial} database, where RT ranks fourth. The \texttt{rel-trial} database contains clinical trial records, which are highly specialized and domain-specific. With only 13,779 task-specific samples and no relational context at hop = 0. Therefore, RT cannot fully demonstrate its capacity, as it is inherently designed to exploit cross-table dependencies.

The classification task that incorporates rows from tables in the immediate neighborhood (hop = 1) exhibits distinct performance rankings. RT outperforms all methods with an average rank of 1.2 followed by Griffin (2.2) and TabPFN 2.5 (3.0). An exception is TabPFN 2.5, which slightly edges out RT on the \texttt{rel-avito user-visits} classification task. The classification tasks in the hop = 2 experiment demonstrate the superiority of the RT method over other baselines, with an average rank of 1.0. Griffin is the second best (2.4), followed by DBformer and LightGBM. TabPFN 2.5 delivers the weakest results, underscoring that single-table learning methods fall short when tackling relational databases with rich, complex referential integrity. The superiority of RT is evident when we consider fifteen experimental cases that make up three hops and five datasets. Figure ~\ref{fig:win_matrix_classification} shows pairwise model comparisons where RT wins against LightGBM (14/15), TabPFN 2.5 (13/15), and Griffin (14/15). Griffin is the second most competitive method overall, winning in 11/15 cases against LightGBM and 12/15 against DBFormer. TabPFN 2.5 outperforms RT in two cases. The first is the \texttt{rel-trial study-outcome} task at hop = 0, where the small number of task-specific samples (13,779) favors TabPFN 2.5, which is specifically designed for small datasets with fewer than 10,000 samples~\citep{ye2025tabpfnlimitation}. The second is the \texttt{rel-avito user-visits} task at hop = 1, where the performance gap is negligible (TabPFN 2.5: 64.5 vs RT: 63.9), suggesting that this may reflect random variance rather than a systematic advantage.

\subsection {Regression performance}

The single table regression task (hop = 0) reveals that RT is the best method (average rank 2.0), followed by Griffin (2.8) and TabPFN 2.5 (3.1). Specifically, RT ranks 3rd and 4th on \texttt{rel-avito ad-ctr} and \texttt{rel-stack post-vote} regression tasks, respectively. Despite the worst rank, DBformer results in the best RMSE scores on the \texttt{rel-avito ad-ctr} regression task when RT ranks 3rd. Regression tasks that incorporate rows from adjacent tables (hop = 1) show a shift in ranking. TABPFN 2.5 matches RT in performance, with both methods achieving an average rank of 2.4. DBFormer (3.6) performs better than LightGBM (3.8) in the hop = 1 setting. In the hop = 2 configuration, RT gains the best average rank of 2.0, while TabPFN 2.5 (2.6) and Griffin (2.8) follow. The win matrix in Fig.~\ref{fig:win_matrix_regression} summarizes pairwise model comparisons across 15 regression tasks, three hops, and five databases. RT wins on 9/15 tasks against LightGBM and DBFormer and on 12/15  against TabPFN 2.5 and Griffin. TabPFN 2.5 shows a notably stronger performance in regression than in classification, outperforming LightGBM and DBFormer on 11/15 tasks. Interestingly, TabPFN 2.5 outperforms RT on the \texttt{rel-stack post-votes} regression task for all three hop settings. TabPFN 2.5, which was pretrained on 130 million synthetic datasets~\citep{hollmann2025tabpfn}, may have seen comparable distributional structures during this pretraining, which could give TabPFN 2.5 an advantage over this specific regression task.


\subsection {Effects of relational hops}

Intuitively, hop = 2 should outperform hop = 1, since it pulls in richer contextual row samples from related tables via referential links. However, all methods, except RT, perform worse under hop = 2 than hop = 1 in the \texttt{rel-trial study-outcome} classification task. The unexpected underperformance of hop=2 may be due to the uniqueness in how health records data are spread across relational tables within limited samples. RT is the only model that consistently improves across all hop settings on \texttt{rel-trial}, indicating its attention mechanism better exploits deeper relational context even with limited samples.

The regression results exhibit a pattern comparable to the classification results. The results of the hop=2 setting do not consistently outperform those of the hop=1 setting. The performance gap between these two hop configurations remains small for relational models (DBFormer, Griffin, and RT). For example, Griffin's RMSE remains unchanged in \texttt{rel-avito} (0.82 $\rightarrow$ 0.82) and \texttt{rel-f1} (0.68 $\rightarrow$ 0.68) from hop = 1 to hop = 2. Similarly, DBFormer shows no improvement in \texttt{rel-stack} (0.36 $\rightarrow$ 0.36), and RT's improvement in \texttt{rel-amazon} is minimal (0.47$\rightarrow$
0.45). In many cases, the improvement from hop = 1 to hop = 2 is minimal, raising the question of whether the performance gain justifies the high computational cost, as evidenced by the training times reported in Table~\ref{tab:training_time_rel_f1_driver_dnf}. For example, Griffin's training time increases from 202 minutes at hop = 1 to 403 minutes at hop = 2 on the \texttt{rel-f1 driver-dnf} task, a 2$\times$ increase in training cost, while the AUC improvement is only 2\%.

\begin{table}[t]
\centering
\caption{Computational time comparison of the deep RDB methods on the 
\emph{rel-f1} \emph{driver-dnf} task across different hop settings.}
\label{tab:training_time_rel_f1_driver_dnf}
\begin{tabular}{ccccccc}
\toprule
& \multicolumn{2}{c}{DBFormer} 
& \multicolumn{2}{c}{Griffin} 
& \multicolumn{2}{c}{RT} \\
\cmidrule(lr){2-3} \cmidrule(lr){4-5} \cmidrule(lr){6-7}
Hop 
& \begin{tabular}[c]{@{}c@{}}Train\\(min)\end{tabular} 
& \begin{tabular}[c]{@{}c@{}}Test\\(sec)\end{tabular}
& \begin{tabular}[c]{@{}c@{}}Train\\(min)\end{tabular} 
& \begin{tabular}[c]{@{}c@{}}Test\\(sec)\end{tabular}
& \begin{tabular}[c]{@{}c@{}}Train\\(min)\end{tabular} 
& \begin{tabular}[c]{@{}c@{}}Test\\(sec)\end{tabular} \\
\midrule
0 & 11 & 1 & 14  & 1 & 101 & 2 \\
1 & 17 & 1 & 202 & 3 & 137 & 2 \\
2 & 23 & 1 & 403 & 7 & 158 & 2 \\
\bottomrule
\end{tabular}
\end{table}



\section{Conclusions}

This paper presents one of the first systematic benchmarking studies of state-of-the-art deep learning methods for relational databases. Using a unified evaluation protocol with consistent data splits, computational budgets, and standardized metrics, we demonstrate that the RT method is the strongest in both classification and regression tasks. In other words, transformer-based methods are superior to conventional graph-based modeling of relational databases. Although RT and Griffin are built for relational data tables connected by primary and foreign keys, they still outperform leading state-of-the-art tabular learning approaches, such as TabPFN 2.5, in single-table (hop = 0) classification and regression benchmarks. The gain from increasing hop from 1 to 2 is often minor relative to the substantial increase in computation, casting doubt on whether expanding relational neighborhoods is worth the cost. Developing relational deep learning methods that can reliably exploit multi-hop relational context with lower computational overhead remains an important open problem for future research.





\bibliographystyle{elsarticle-num}
\bibliography{relational}

\end{document}